\documentclass[aps,prb,twocolumn,showpacs,superscriptaddress,amsmath,amssymb,floatfix,longbibliography,10pt]{revtex4-2}

\usepackage{graphicx}
\usepackage{amssymb}
\usepackage[toc,page]{appendix}
\usepackage{color}
\usepackage{comment}
\usepackage{hyperref}
\usepackage{epsf}
\usepackage{float}
\usepackage{longtable}
\usepackage{booktabs}
\usepackage{threeparttable}

\usepackage{subfigure,amsmath,verbatim,moreverb}
\usepackage{tabularx}
\usepackage{adjustbox}
\usepackage{lipsum}
\usepackage{latexsym}
\usepackage{dcolumn}
\usepackage{amsmath}
\usepackage{dcolumn}

\newcommand{\R}{\mathbf{r}}

\newcommand{\RRef}[1]{Ref. \citenum{#1}}

\newcommand{\Tab}[1]{Tab. \ref{#1}}

\newcommand{\be}{\begin{equation}}
\newcommand{\ee}{\end{equation}}
\newcommand{\bea}{\begin{eqnarray}}
\newcommand{\eea}{\end{eqnarray}}
\newcommand{\bean}{\begin{eqnarray*}}
\newcommand{\eean}{\end{eqnarray*}}

\begin{document}

\title{ Towards adiabatic-connection interpolation model with broader applicability }
\author{Lucian A. Constantin}
\affiliation{Institute for Microelectronics and Microsystems (CNR-IMM), 
Via Monteroni, Campus Unisalento, 73100 Lecce, Italy}
\author{Szymon \'Smiga}
\affiliation{Institute of Physics, Faculty of Physics, Astronomy and Informatics, Nicolaus Copernicus 
University in Toru\'n, c}
\author{Fabio Della Sala}
\email{fabio.dellasala@imm.cnr.it}
\affiliation{Center for Biomolecular Nanotechnologies, Istituto Italiano di
Tecnologia, Via Barsanti 14, 73010 Arnesano (LE), Italy}
\affiliation{Institute for Microelectronics and Microsystems (CNR-IMM), Via
Monteroni, Campus Unisalento, 73100 Lecce, Italy}

\date{\today}

\begin{abstract}
The Adiabatic Connection Integrand Interpolation (ACII) method represents a general path for calculating correlation energies in electronic systems within the Density Functional Theory.
ACII functionals include both exact-exchange and the second-order correlation energy, as well as an interpolating function toward the strictly-correlated electron (SCE) regime. 
Several interpolating functions have been proposed in the last years targeting different properties, yet an accurate ACII approach with broad applicability is still missing.
Recently, we have proposed an ACII functional
that was made accurate for the three-dimensional (3D) uniform electron gas as well as for model metal clusters.
In this work 
we present an ACII functional (named genISI2) which is very accurate for both three-dimensional (3D) and two-dimensional (2D) uniform electron gases and for the quasi-2D infinite barrier model, where most of the exchange-correlation functionals fail badly, as well as for strongly correlated two-electrons systems.
Using the exact-exchange Kohn-Sham orbitals, we have also assessed the genISI2 for various molecular systems, showing a superior performance with respect to the other ACII methods for total energies, atomization energies, and ionization potentials. The genISI2 functional can thus find application in a broad range of systems and properties. 
\end{abstract}

\maketitle

\section{Introduction}
\label{sec1}
Nowadays, the ground-state Kohn-Sham (KS) Density Functional Theory (DFT) \cite{hohenberg1964inhomogeneous,kohn1965self} is widely used in electronic structure calculations of finite and extended systems \cite{burke12,book2,burke12,giustino_book,becke14,jones15}, providing an adequate ratio between the accuracy and computational time. The basic DFT variable is the ground-state electronic density $n(\R)$, which implicitly determines all the ground-state properties of the electronic system \cite{kohn1996density,kohn1999nobel}. In the KS-DFT,  $n(\R)$ is found by solving the Euler equation via one-particle orbitals $\{\phi_{i,\sigma}(\R)\}$, such that the non-interacting kinetic energy functional is treated exactly as $T_s[n(\R)]=\langle\Phi_n^{min}|\hat{T}|\Phi_n^{min}\rangle=\int d\R\;\sum_{i,\sigma}^{occ}|\nabla\phi_{i,\sigma}|^2/2$, where $\Phi_n^{min}$ is the Slater determinant build with KS one-particle orbitals $\{\phi_{i,\sigma}(\R)\}$, that yields the density $n(\R)$ and 
minimizes the expectation value of the kinetic energy operator $\langle\hat{T}\rangle$. Thus, only the exchange-correlation (XC) energy functional $E_{xc}[n(\R)]$ must be approximated. We recall that the XC energy functional should describe the electron-electron interactions beyond the classical Hartree energy $U[n]=(1/2)\int d\R\int d\R'\;n(\R)n(\R')/|\R-\R'|$. 

An elegant (and exact in principle) definition of the XC functional uses the 
adiabatic connection (AC) method: \cite{langreth1975exchange,gunnarsson1976exchange,savin2003adiabatic,
cohen2007assessment,ernzerhof1996construction,burke1997adiabatic,colonna1999correlation}
\begin{eqnarray}
E_{xc}[n]&=&\int_0^1 d\alpha\;W_{xc,\alpha}[n],\nonumber \label{eq:aint}\\ 
W_{xc,\alpha}[n]&=&\langle\Psi_n^{min,\alpha}|\hat{V}_{ee}|\Psi_n^{min,\alpha}]\rangle-U[n],
\label{eq1}
\end{eqnarray}
where $\hat{V}_{ee}$ is the Coulomb repulsion operator, and
$\Psi_n^{min,\alpha}$ is the antisymmetric wave function that yields the density $n(\R)$ and 
minimizes the expectation value $\langle\hat{T}+\alpha \hat{V}_{ee}\rangle$,
with $\alpha\ge 0$ being the coupling constant (also known as interaction strength).

There are many XC functionals constructed in the framework of the AC method, 
including accurate hybrid functionals \cite{burke1997adiabatic,kohn1996density,becke1993density,adamo1998exchange,UPBH} 
and the most sophisticated fifth-rung functionals \cite{Jacob_ladder}(which uses the unoccupied orbitals in the functional definition) such as the random phase approximation (RPA) \cite{hess11,Eshuis2012,chen17} and double-hybrid(DH) functionals \cite{goerigk2014double}.

In the fifth-rung are also included the functionals investigated in this work, i.e., the one based on  a $W_{xc,\alpha}[n]$  model \cite{seidl1999strictly,seidl2000simulation,seidl2000density,
perdew2001exploring,magyar2003accurate,
gori2009electronic,liu2009adiabatic,liu2009adiabatic2,
sun2009extension,
seidl2010adiabatic,gori2010density,mirtschink2012energy,zhou2015construction,
vuckovic2016exchange,fabiano2016interaction,vuckovic2017interpolated,vuckovic2017simple,
giarrusso2018assessment,
vuckovic2018restoring,kooi2018local,constantin2019correlation,fabiano2019investigation,
smiga2020modified,smiga2022selfconsistent,
vuckovic2023density,constantin2023adiabatic,jana2023semilocal},
which interpolates between the weak- ($\alpha\rightarrow 0$) and strong- ($\alpha\rightarrow \infty$) interaction limits, and that we refer to as Adiabatic Connection Integrand Interpolation (ACII).

We recall that the asymptotic behaviors of $W_{xc,\alpha}[n]$ are known exactly \cite{seidl2000simulation,gorling1993correlation,jana2020generalizing,liu2009adiabatic}:
\begin{eqnarray}
W_{xc,\alpha\rightarrow 0}[n]&=&W_0[n]+W'_0[n]\alpha+ \dots \nonumber \\ &+& W^{(m)}_0[n]\alpha^m+\dots,  \label{eq:limalpha0} \\
W_{xc,\alpha\rightarrow\infty}[n]&=&W_\infty[n]+W'_\infty[n]\alpha^{-1/2} + \nonumber \\
&+& W^{(2)}_\infty[n]\alpha^{-3/2}+\dots`,
\label{eq2}
\end{eqnarray}
Here, $W_0[n]=E_x[n]$ is the exact DFT exchange functional,
$W'_0[n]=2E_c^{GL2}[n]$, and $W^{(m)}_0[n]=(m+1)E_c^{GL_{m+1}}[n]$, where GL indicated the G\"{o}rling-Levy perturbation theory
\cite{gorling1994exact,gorling1993correlation,gorling1995hardness}.
Usually, only the GL2 correlation ($E_c^{GL2}$) is included in the ACII functionals. In some schemes, the GL2 terms are approximated with a semilocal functional \cite{cohen2007assessment,smiga2020modified,constantin2023adiabatic,jana2023semilocal}: in this way, however, the whole functional does not belong anymore to the fifth-rung.

In the strong-interaction limits the strictly correlated electron (SCE) approach \cite{gori2009density,malet2012strong,friesecke2022strong} becomes exact. On the other hand, the SCE method becomes computationally very demanding, especially for larger systems, such that usually $W_\infty[n]$ and $W'_\infty[n]$ are
approximated using generalized gradient approximation (GGA) or meta-GGA formulas \cite{seidl2000density,constantin2019correlation,smiga2022selfconsistent,jana2023semilocal}.

One of the first and most known ACII functional is Interaction Strength Interpolation (ISI)\cite{seidl2000simulation}. In the last years, different approaches and investigations have been presented\cite{fabiano2016interaction,
giarrusso2018assessment,
vuckovic2018restoring,constantin2019correlation,fabiano2019investigation,
smiga2020modified,smiga2022selfconsistent,constantin2023adiabatic}, including those based on the Hartree-Fock (HF) density
\cite{seidl2018communication,daas20,daas2021noncovalent,daas22,daas2023regularized}.
Efficient implementation of ACII methods is also available in public quantum-chemistry codes \cite{turbo2023}.


After the $\alpha$-integration in Eq. \eqref{eq:aint}, the total XC energy of ACII functionals becomes a non-linear function (${\cal F}$) of the $E_x$,$E^{GL2}_c$,  $W_\infty$ and $W'_\infty$ ingredients:
\begin{equation}
E_{xc}[n]={\cal F}(E_x,E^{GL2}_c,W_\infty,W'_\infty) \label{eq:eexc} \; .
\end{equation}
Eq. \eqref{eq:eexc} thus resembles DH functionals \cite{goerigk2014double}: however, ACII  employs a non-linear dependence from the GL2 term and does not diverge for systems with vanishing gaps \cite{fabiano2016interaction,smiga2020modified,turbo2023}, which is a clear superiority with respect to DH approaches.

Recently, we have proposed the UEG-ISI functional \cite{constantin2023adiabatic}, an ACII functional (valid in the limit of $E_c^{GL2}\rightarrow-\infty$) that gives an improved description of the three-dimensional (3D) uniform electron gas (UEG).
Then, the UEG-ISI has been supplemented by a term that depends on  $E_c^{GL2}$, finally yielding the genISI functional \cite{constantin2023adiabatic}, with good performance for jellium clusters and atoms.

In this work, we construct the genISI2 XC functional as a revision of
the one proposed in Ref. \onlinecite{constantin2023adiabatic},
fulfilling the negativity constraint of the correlation energy ($E_c\le 0$) for any possible values of the ingredients $W_0$, $W'_0$, $W_\infty$ and $W'_\infty$. 
Both the genISI and genISI2 functionals recover UEG-ISI in the limit of $E_c^{GL2}\rightarrow-\infty$.

We assess the genISI and genISI2 functionals for various systems such as the two-dimensional (2D) UEG, the quasi-2D infinite barrier model (IBM), the Hooke's atom and the dissociation of the spin-restricted $H_2$ molecule, which are very hard tests for XC functionals:

The 2D UEG is a paradigm for 2D electronic systems \cite{attaccalite2002correlation}, and the 2D local density approximation (LDA) XC functional yields already quite a good accuracy for the total energy of many 2D systems \cite{2DLDA1}. However, most of 3D XC functionals fail badly for 2D UEG, and only few high-level methods, such as the inhomogeneous Singwi-Tosi-Land-Sj\H{o}lander (ISTLS) can accurately describe the 2D UEG \cite{ISTLS1,ISTLS2}. 

The quasi-2D IBM \cite{pollack2000evaluating,chiodo2012nonuniform,karimi2014three,
constantin2016simple,kaplan2018collapse,horowitz2023construction}, which is a severe test for the dimensional crossover from 3D to 2D, is poorly described both at the RPA and the semilocal (GGA and meta-GGA) levels of theory, e.g. see Refs. \onlinecite{constantin2019correlation,constantin2016simple}. For example, the popular Perdew–Burke–Ernzerhof (PBE) GGA \cite{perdew1996generalized} and the Tao-Perdew-Staroverov-Scuseria (TPSS) \cite{TPSS} meta-GGA XC functionals fail badly for this test.  

The Hooke's atom \cite{taut1994two,coe2008entanglement,ludena2004exact,liang2011hooke,o2003wave,gori2009study,lam1998virial,burke1998unambiguous}, also named Harmonium, represents two interacting electrons in an isotropic harmonic potential of frequency $\omega$ ($\omega=\sqrt{\kappa}$, where $\kappa$ is the force constant). At small values of $\omega$, the electrons are strongly correlated, and at large frequencies, they are tightly bound \cite{gill2005electron}. Because ACII formulas interpolate between the weak- and strong-correlation limits, this 
is a very important test.

The $H_2$ dissociation curve using a spin-restricted formalism, which is a widely studied  prototype of a strongly correlated system 
\cite{cohen08,Mori-Sanchez.Cohen.ea:Discontinuous.2009,Cohen.Mori-Sanchez.ea:Second-Order.2009,fabiano2016interaction,Janesko.Proynov.ea:Practical.2017,fuchs2005desc,BGE2,vuckovic2016XC,vuckovic2017simple,peach2008adiabatic,teale2010accurate}. While most of  XC functionals gives the correct energy at the
equilibrium distance, only the most advanced functionals can correctly describe the static-correlation
at infinite distance. \cite{Janesko.Proynov.ea:Practical.2017,BGE2,vuckovic2017simple,zhang21,science21}

In addition, we considered the more conventional molecular systems, investigating the accuracy for total energies, atomization energies and ionization potentials.
The performances of genISI and genISI2 are then discussed in comparison with the other well-known ACII functionals.

The paper is organized as follows:
In section \ref{sec2}, we overview the genISI and 
present the construction of genISI2 XC functional;
Computational details are described in Section \ref{seccd}, while in section \ref{sec4}, we report the results for various systems. 
 Finally, in section \ref{secconc}, we summarize our conclusions. 

\section{Theory}
\label{sec2}
\subsection{Overview of the genISI exchange-correlation functional}
\label{sec21}

In the UEG-ISI functional\cite{constantin2023adiabatic} the AC integrand is:
\begin{equation}
W^{UEG-ISI}_{xc,\alpha}[n]=W_\infty[n]+\frac{b(2+c\alpha+2d \sqrt{1+c\alpha})}
{2\sqrt{1+c\alpha}(d+\sqrt{1+c\alpha})^2},
\label{eq6}
\end{equation}
where
\begin{eqnarray}
&& b=b[n]=(W_0[n]-W_\infty[n])(1+d) \ge 0 , \nonumber\\
&& c=c[n]=b[n]^2 / ( 4 W'^2_\infty[n] ) \ge 0 ,\nonumber \\
&& d=3.5.
\label{eq7}
\end{eqnarray}

The parameter $d$ has been optimized, minimizing the error for the correlation energy per particle ($\epsilon_c$) of the 3D UEG.
For small $\alpha$ we have:
\begin{equation}
W^{UEG-ISI}_{xc,\alpha}[n] \rightarrow W_0[n] -s[n]\alpha + \ldots
\label{eq:uegisismall}
\end{equation}
with
\begin{equation}
s[n]=\frac{1+d}{4}
\frac{(W_0[n]-W_{\infty}[n])^3}{W'^2_{\infty}[n]} \ge 0.
\end{equation}
Note that there are no exact conditions for the coefficient $s[n]$, as the UEG-ISI functional is built for metallic systems, in which $E_c^{GL2}=-\infty$, see the discussion
in Ref. \onlinecite{constantin2023adiabatic}. 
The UEG-ISI XC energy is:
\begin{eqnarray}
E^{UEG-ISI}_{xc}[n]&=&\int^1_0 d\alpha\; 
W^{UEG-ISI}_{xc,\alpha}[n] \nonumber\\
&=&W_\infty[n]+\frac{b[n]}{d+\sqrt{1+c[n]}}.
\label{eq11}
\end{eqnarray}

The genISI functional adds a $E_c^{GL2}$ dependent term to UEG-ISI; it has the following expression for the AC integrand:
\begin{eqnarray}
W^{genISI}_{xc,\alpha}[n]&=&W^{UEG-ISI}_{xc,\alpha}[n]+\nonumber\\
&& \frac{A[n]\alpha}{(1+m\, r[n]p[n]\alpha)^3},
\label{eq12}
\end{eqnarray}
with 
\begin{eqnarray}
\nonumber\\
p[n]&=&W'_0[n]/W_0[n] \ge 0 , \nonumber\\
A[n]&=& W'_0[n] +s[n], \nonumber \\
r[n]&=& \Big( \frac{W_0[n]}{W_\infty[n]}\Big)^3 \ge 0,\nonumber \\
m&=& 18.0.
\label{eq13}
\end{eqnarray}

The parameter $m$ has been obtained by fitting to the correlation energy of Harmonium with force constant $\kappa=1/4$\cite{taut1994two,constantin2019correlation}. The genISI functional has been constructed so that for small $\alpha$ it recovers the exact expansion:
\begin{equation}
W_{xc,\alpha}[n] \rightarrow  W_0[n] +(2 E_c^{GL2})\alpha + \ldots
\label{eq:wxcsmall}
\end{equation}
Integrating over $\alpha$, we obtain a simple analytical expression for the genISI XC energy:
\begin{eqnarray}
&& E^{genISI}_{xc}[n] =
\nonumber\\
&& = E^{UEG-ISI}_{xc}[n]+\frac{A[n]}{2(m\, r[n]p[n]+1)^2}.
\label{eq15}
\end{eqnarray}
%

%

%

%

\subsection{The genISI2 functional}
\label{sec22}
For the genISI2 functional, we consider the following expression:
\begin{eqnarray}
&& W_{xc,\alpha}^{genISI2}=W_{xc,\alpha}^{UEG-ISI}+W_{xc,\alpha}^{a_1}+W_{xc,\alpha}^{a_2},\nonumber\\
&& W_{xc,\alpha}^{a_1}=\frac{W'_0[n]\alpha}{(1+l_1 r[n]p[n]\alpha)^3} \le 0,\nonumber\\
&& W_{xc,\alpha}^{a_2}=\frac{ -W_{xc,\alpha}^{UEG-ISI}[n]+W_0[n]}{(1+l_2 r[n]p[n]\alpha)^3} \ge 0,\nonumber\\
\label{eq15tt}
\end{eqnarray}.

For small $\alpha$ the genISI2 behaves as:
\begin{eqnarray}
 && W_{xc,\alpha}^{genISI2} \rightarrow  W_0[n] +  \nonumber \\
 && + 
 \left( 
  2+\frac{6 l_2 W_0[n]^2 (W_{xc,\alpha=0}^{UEG-ISI}[n]-W_0[n])) }{W_\infty[n]^3}
 \right) \alpha E_c^{GL2}+... \nonumber \\
 &&
\end{eqnarray}
Thus, using  Eq. \eqref{eq:uegisismall}, we have that genISI2 recovers the exact limit in Eq. \eqref{eq:wxcsmall}.

The genISI2 XC energy
\begin{equation}
E^{genISI2}_{xc}[n] = \int^1_0 d\alpha\;
W^{genISI2}_{xc,\alpha}[n] ,
\label{eq15t3}
\end{equation}
does not have a simple analytical expression (as genISI). Still, the integral over the coupling constant can be numerically computed with high efficiency and accuracy (e.g., using a Gaussian quadrature with only 16 points). 

The genISI2 functional recovers all the exact conditions fulfilled by revISI \cite{gori2009electronic}, LB \cite{liu2009adiabatic}, genISI, and additionally, the genISI2 correlation energy is always negative. This is not the case for the genISI functional. We mention that for all the studied systems, we have found the correct behavior of genISI (i.e., $E_c^{genISI}\le 0$), however in principle, $E_c^{genISI}$ can be positive, probably outside of the physical range of $W_0$, $W'_0$, $W_\infty$ and $W'_\infty$ ingredients.
In fact, the quantity $A[n]$ can be positive for small values of $E_c^{GL2}$ (see Appendix for further details.)

Moreover, by construction, 
\begin{equation}
\lim_{W'_0\rightarrow 0} E_c^{genISI2}=0,
\label{eq15t3d}
\end{equation}
a condition not satisfied by genISI, see Eq. \eqref{eq15}. We recall that $W'_0$ vanishes not only for any one-electron systems (where genISI is also exact because of $W_0=W_\infty$),  but also for a perfect insulator (where genISI will fail), an interesting model system used in the DFT and solid-state physics development \cite{kohn1986discontinuity,kohn1957effective}. Note that the correlation energy of the jellium-with-gap model vanishes in the limit of infinite band gap energy \cite{rey1998virtual,fabiano2014generalized}, such that Eq. \eqref{eq15t3d} remains valid in this case.

 Finally, the parameters  $l_1$ and $l_2$ were found from the minimization of the mean absolute relative error (MARE) of several small spherical systems, as described in the following subsection.

\subsection{Small systems with accurate ingredients}
\label{sec23}
For few spherical systems (Harmonium with force constant $\kappa = 1/4$, the two-electron exponential density [$n(r) = 2 \exp({-2r})/\pi$ ], and the He, Be, Ne), the ingredients 
(
$W_0$, 
$W'_0$, 
$W^{SCE}_\infty$, 
and 
$W'^{SCE}_\infty$
)
are known (almost) exactly from literature (see Table \ref{ta1}). 
For the Neon atom, we recomputed  $W_0$ and $W'_0$ using the 
the coupled-cluster single double and perturbative triple [CCSD(T)]
density and inversion technique (see Sect. \ref{seccd} for details).

%
\begin{figure}[hbt]
\includegraphics[width=\columnwidth]{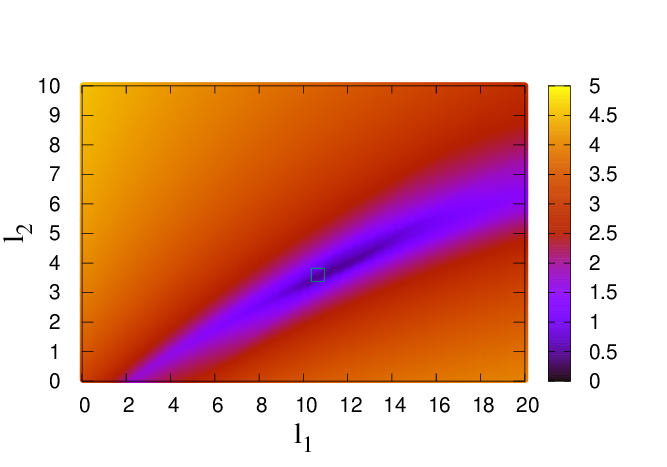}
\caption{$\log [$ MARE($l_1,l_2$)] (in \%) of correlation energies computed for the systems of Table \ref{ta1}, from the genISI2 expression of Eq. \eqref{eq15t3}, as a function of the $l_1$ and $l_2$ parameters. The minimum of the $\log (MARE(l_1,l_2))$ is at $l_1=10.65$ and $l_2=3.6$ (shown with an empty square), i.e. the parameters used in the genISI2 functional.
}
\label{f1bb}
\end{figure}
%
Fig. \ref{f1bb} reports the MARE($l_1,l_2$) of the correlation energies of these systems, computed with the genISI2 expression, as a function of the $l_1$ and $l_2$ parameters. The minimum error is found at $l_1=10.65$ and $l_2=3.6$, which we use to define
the genISI2 functional.

In Tab. \ref{ta1}, we show the correlation energies of these systems, computed from several ACII functionals. The genISI2 functional gives the best MARE ($\approx 2\%$) showing that the genISI2 functional form is flexible enough to describe the correlation in these systems
%
\begin{table}[htbp]
\caption{Exact (or almost exact) ACII ingredients (in Ha) and negative of correlations energies ($-E_c$ in mHa) from several ACII functionals, computed for the Harmonium with force constant $\kappa = 1/4$, the two-electron exponential density [$n(r) = 2 exp{(-2r)}/\pi$, Exp.], and the He, Be, and Ne atoms. The exact values are from Ref. \onlinecite{constantin2023adiabatic} and references therein.  
In the last column, we show the MARE (in \%) of each method. 
Best ACII results are in boldface. 
}
\begin{tabular}{ccccccc}
\hline
 & Harm.& Exp. & He & Be & Ne \\ \hline \hline 
$W^\text{SCE}_\infty$ 
& -0.743$^a$ & -0.910$^c$ & -1.500$^c$ & -4.021$^h$ & -20.035$^h$ \\ 
$W^{'\text{SCE}}_\infty$ 
& 0.208$^a$ & 0.293$^e$ & 0.621$^g$ & 2.590$^g$ & 22.0$^g$ \\ 
$W_0$ 
& -0.515$^b$ & -0.625$^c$ & -1.024$^f$ & -2.673$^f$ & -12.078$^i$\\ 
$W^{'}_0$ 
&-0.101$^b$ & -0.093$^d$ & -0.095$^g$ & -0.246$^f$ & -0.948$^i$\\ 
& & &  &  & & MARE (\%) \\ \hline
SPL 
& 35.9 & 35.6 & 39.9 & 104.0 & 428.8 & 7.1 \\ 
LB 
& {\bf 38.5} & 37.8 & 41.6 & 108.1 & 436.8 & 5.8 \\ 
ISI 
& 36.6 & 36.4 & 40.5 & 102.4 & 414.3& 5.1 \\ 
revISI 
& 37.0 & 36.9 & 40.8 & 101.7 & 409.3 & 4.1 \\ 
genISI 
& 39.6 & {\bf 37.4} & 39.3 & 106.5 & 415.7 & 5.8 \\ 
{genISI2} 
& {37.2} & {38.0} & {\bf 42.3} & {\bf 97.2} & {\bf 391.9} & {\bf 1.8} \\ 
\hline
Exact 
& 38.5 & 37.3 & 42.1 & 94.4 & 391.0 & - \\ 
\hline 
\hline
\end{tabular} \\
$a)$ from Ref. \onlinecite{liu2009adiabatic}, 
$b)$ from Tab IV of Ref. \onlinecite{magyar2003accurate}, 
$c)$ from Tab. I of Ref. \onlinecite{seidl1999strictly},
$d)$ from Tab. I of Ref. \onlinecite{iva98},
$e)$ Ref. \onlinecite{privcomm},
$f)$ for Tab. II and Tab. IV of Ref. \onlinecite{colonna1999correlation}, using accurate density,
$g)$ from Tab. 1 of Ref. \onlinecite{gori2009electronic},
$h)$ from Tab. I of Ref. \onlinecite{seidl07}, using accurate density,
$i)$ computed in this work at the CCSD(T)/unc-aug-cc-pV6Z density.
\label{ta1}
\end{table}
%

%
\begin{figure}[hbt]
\includegraphics[width=\columnwidth]{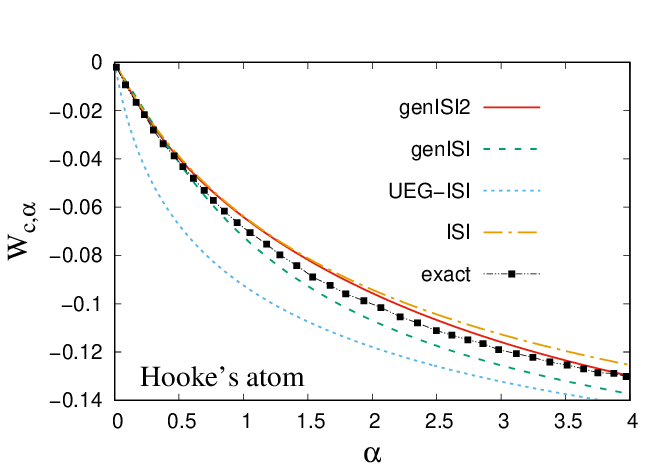}
\includegraphics[width=\columnwidth]{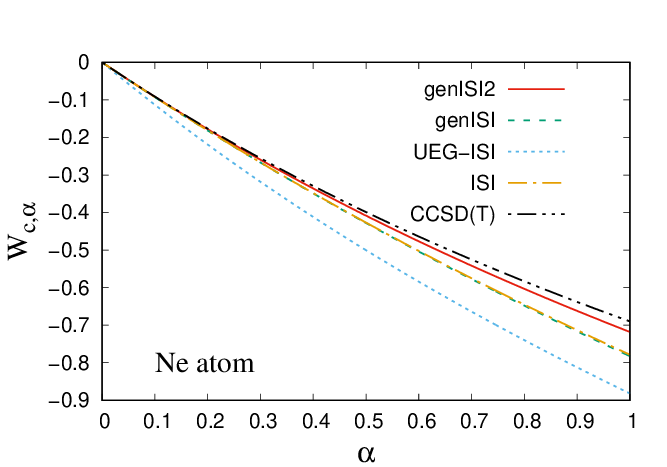}
\caption{The adiabatic connection correlation integrand $W_{c,\alpha}=W_{xc,\alpha}-W_0$, for Harmonium ($\kappa=1/4$) (upper panel) and for the Ne atom (lower panel). The Harmonium exact curve is from Ref. \onlinecite{magyar2003accurate} and the Ne atom CCSD(T) reference (with uncontracted aug-cc-pCVQZ basis set) is from Ref. \onlinecite{teale2010accurate}. We use the exact/accurate (SCE) ingredients shown in Table \ref{ta1}.   
}
\label{f1a}
\end{figure}
%
In Fig. \ref{f1a}, we show a comparison between UEG-ISI, ISI, genISI, and genISI2 for adiabatic connection correlation integrand $W_{c,\alpha}=W_{xc,\alpha}-W_0$ of Harmonium ($\kappa=1/4$) and Ne atom. We observe that both genISI and genISI2 curves are smooth and realistic, being close to the exact result. On the other hand, the UEG-ISI integrand does not have the right slope at $\alpha=0$, showing a strong underestimation. We recall that the area under the curves for $0\le \alpha \le 1$ represents the correlation energy $E_c=\int_0^1 d \alpha W_{c,\alpha}$. However, we mention that for both atoms, the genISI2 gives the best performance, being better than ISI and genISI. Noting that the CCSD(T) correlation energy of the reference curve of the lower panel of Fig. \ref{f1a} is about 14 mHa bigger than the benchmark reference (see Table \ref{ta1}), we may even expect that $W_{\alpha,c}^{genISI2}$ is the most accurate shown result for the Ne atom. 

\section{Computational details:}
\label{seccd}
\subsection{ (quasi-)2D systems}
\label{seccd1}
The calculations for the quasi-2D IBM  in Sect. \ref{sec43} use exact orbitals and densities \cite{pollack2000evaluating}. Thus $W_0$ and $W'_0$ ingredients are exactly computed \cite{betbeder1996quasi}, while for the strong-interaction limit ingredients $W_\infty$ and $W'_\infty$ we use (at any quantum well thickness) accurate interpolations between the 2D and 3D limits, as explained in detail in Sec. \ref{sec43}.
Note that for $W_\infty(L_{max})$ and  $W'_\infty(L_{max})$ we use
the meta-GGA functional of Ref. \onlinecite{perdew2004meta} and  Ref.  \onlinecite{seidl2000density}, respectively.
For the 2D uniform electron gas in Sect. \ref{sec35l}, we use the exact ingredients.

\subsection{ Atoms and Molecules}
\label{seccd2}
All calculations have been performed with a locally modified version of {\tt ACES II}~\cite{acesII} program. All ACII results have been obtained in a post-self-consistent-field (SCF) fashion, using as a reference OEPx SCF converged quantities (i.e. orbitals, orbital energies, and densities). As in our previous studies \cite{smiga2016self,smiga2019self,smiga2020modified,smiga2022selfconsistent} to solve OEPx equations, we have employed the finite-basis set procedure from \RRef{ivanov:1999:OEP}. 
The hPC model \cite{smiga2022selfconsistent} was used in all calculations to evaluate the $W_\infty$ and $W'_\infty$.

 In particular, we have investigated:

 \begin{itemize}

 \item[-] \textbf{Harmonium:} We have performed calculations for various values of $\omega $ ($\in 0.03 \div 1000$) in the Harmonium model\cite{PhysRev.128.2687} using the computational setup and an even-tempered Gaussian basis set (up to f angular momentum functions) from \RRef{B926389F}. 
 
   \item[-] \textbf{Total energies:} the total energies (reported in  \Tab{taTE}) have been calculated for the systems listed in Table I of Ref. \onlinecite{grabowski2014orbital}, using an identical computational setup. 
   
   \item[-] \textbf{Atomization energies:}  AE6 \cite{lynch2003small,haunschild2012theoretical} atomization energies listed in \Tab{taAE6}. These calculations have been performed using uncontracted cc-pVTZ basis sets of Dunning~\cite{dunning:1989:bas} without counterpoise corrections for basis set superposition error (BSSE). The results have been corrected for size-consistency error according to \RRef{vuckovic2018restoring} (Note, however, that this correction is very small for atomization energies).

    \item[-] \textbf{Vertical ionization potentials:} 32 vertical ionization potentials (VIP)\cite{SmigaIP} computed as the energy difference between the neutral and the ionic species\cite{IPszs}. The computational setup, namely basis sets and geometries (in the case of molecules), is identical as in \RRef{SmigaIP}.

 \end{itemize}
For total energies, atomization energies and ionization potentials, we used, as reference data, the CCSD(T) \cite{Raghavachari1989479} results obtained in the same basis set to make a comparison on the same footing and to reduce basis set related errors. 

 \subsection{Accurate reference for atomic systems}
 \label{seccd3}
To compute the reference $W_0$, $W_\infty$, $W'_\infty$, and $E_c^{GL2}$ values reported in Tab. \ref{ta1} and \ref{tab:reforb}, one needs to obtain accurate KS occupied and unoccupied orbitals from which $E_c^{GL2}$ can be evaluated. To this end, we have considered the densities from Full Configuration Interaction (FCI), in the case of Harmonium and He atoms, and CCSD(T), for Ne atom. The KS potential is then obtained using the Wu-Yang (WY)\cite{WY2003} inversion procedure. 
All calculations have been performed in a locally modified PySCF \cite{pyscf} program together with {\it KS-pies} package \cite{ks-pies} for the WY method. In all calculation, the lambda regularization parameter\cite{l_curve_1} was set to $10^{-5}$ and the Fermi-Amaldi guide potential was used to correct the asymptotic behavior of the XC potentials. We used the uncontracted aug-cc-pV6Z \cite{mourik1999a} basis set for He and Ne atoms and for Harmonium the basis set from \RRef{B926389F}.

\section{Results}
\label{sec4}
In this section, we validate the accuracy of the genISI2 functional for the quasi-2D IBM model (see Sect. \ref{sec43}) for the two-dimensional electron gas (see Sect. \ref{sec35l}) 
and for finite systems (see Sect. \ref{SEC:finite}).

\subsection{quasi-2D IBM}
\label{sec43}

Let us consider the quasi-2D IBM quantum well
of thickness $L$ in the $z$-direction
\cite{pollack2000evaluating,chiodo2012nonuniform,karimi2014three,
constantin2016simple,kaplan2018collapse,horowitz2023construction}.
The true 2D uniform electron gas limit is recovered by shrinking
the $z$-coordinate, keeping fixed the total number of electrons
per unit area ($n^{2D}$).
The quasi-2D regime is obtained when
$L \leq \sqrt{3/2}\pi r_s^{2D}=L_{\rm{max}}$ \cite{pollack2000evaluating},
being equivalent to a non-uniform
scaling in one dimension (i.e. $n^z_{\lambda}(x,y,z)=\lambda n(x,y,\lambda z)$, with
$\lambda=L_{max}/L$). The 3D density of this quasi-2D system is
\begin{equation}
n(z)=\frac{2}{L\pi (r_s^{2D})^2}\sin^2(\pi z/L).
\label{eq9}
\end{equation}

We perform similar calculations to those reported in Fig. 6 of Ref. \onlinecite{pollack2000evaluating}.
Thus, we use the following ingredients for the ACII methods:
\begin{equation}
W_0(L)/N=E_x(L)/N,
\label{eql1}
\end{equation}
where $E_x(L)/N$ is the exact exchange per particle calculated with exact orbitals for the quasi-2D IBM of 
thickness $L\le 
L_{max}$. We recall that most semilocal exchange functionals fail badly for the quasi-2D IBM system 
\cite{chiodo2012nonuniform,horowitz2023construction}, diverging in the limit $L\rightarrow 0$. Note that 
\begin{equation}
W_0(0)/N=E_x(0)/N=-4\frac{\sqrt{2}}{3\pi}\frac{1}{r_s^{2D}},
\label{eql2}
\end{equation}
is the exchange energy per particle of the 2D uniform electron gas. 

For $W_\infty(L)/N$ we use the physically motivated interpolation proposed in Ref. \onlinecite{pollack2000evaluating}:
\begin{equation}
\frac{W_\infty(L)}{N}=\Big[ \frac{W_\infty(0)}{W_0(0)}(1-\frac{L}{L_{max}})+
\frac{W_\infty(L_{max})}{W_0(L_{max})}\frac{L}{L_{max}}  \Big] \frac{W_0(L)}{N},
\label{eql3}
\end{equation}
where
\begin{equation}
W_\infty(0)/N=W_\infty^{2D}/N=\Big( \frac{8}{3\pi}-2 \Big)\frac{1}{r_s^{2D}},
\label{eql4}
\end{equation}
and $W_\infty(L_{max})/N$ has been computed using the $W_\infty^{TPSS}$ of Eq. (37)
of Ref.  \onlinecite{perdew2004meta} that is one the most
accurate models for $W_\infty$ of 3D electronic systems.

On the other hand, for $W'_\infty(L)/N$ we use the simplest interpolation
\begin{equation}
\frac{W'_\infty(L)}{N}=\frac{W'_\infty(0)}{N}(1-\frac{L}{L_{max}})+\frac{W'_\infty(L_{max})}{N}\frac{L}{L_{max}}, 
\label{eql5}
\end{equation}
where 
\begin{equation}
W'_\infty(0)/N=W^{'2D}_\infty/N=\frac{1}{2 (r_s^{2D})^{3/2}},
\label{eql6}
\end{equation}
and $W'_\infty(L_{max})/N$ has been computed using $W'^{MGGA}_\infty$ of Eq. (D16) of Ref. 
\onlinecite{seidl2000density}.

Finally, the $W'_0(L)/N$ is known exactly
\begin{equation}
W'_0(L)/N=2 E_c^{GL2}(L)/N,
\label{eql7}
\end{equation}
where $E_c^{GL2}(L)/N$ is given by Eq. (45) of Ref. \onlinecite{betbeder1996quasi}. We observe that for 2D UEG, 
$E_c^{GL2}(0)/N=-0.1925$ Ha, being independent on $r_s^{2D}$ and finite, in contrast to the 3D UEG case where 
$E_c^{GL2}/N \rightarrow -\infty$. In fact, in the high-density limit of 2D UEG ($r_s^{2D}\rightarrow 0$), $\epsilon_c \rightarrow 
E_c^{GL2}(0)/N=-0.1925$ Ha \cite{attaccalite2002correlation}.  

%
\begin{figure}[!hbt]
\includegraphics[width=\columnwidth]{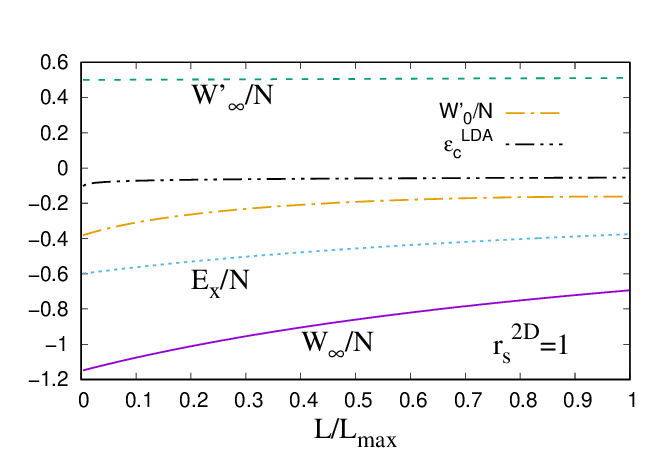}
\includegraphics[width=\columnwidth]{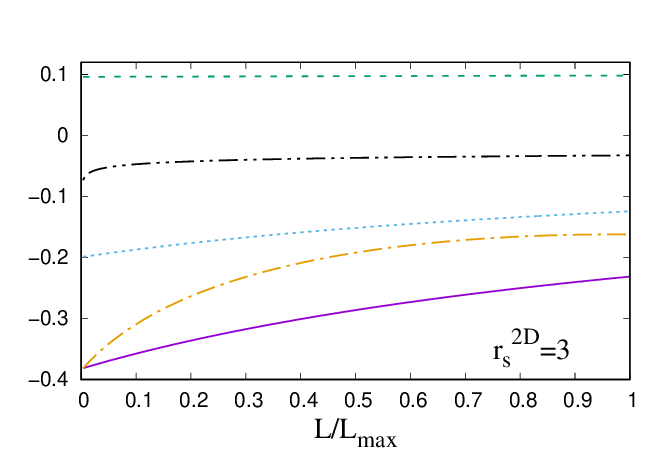}
\includegraphics[width=\columnwidth]{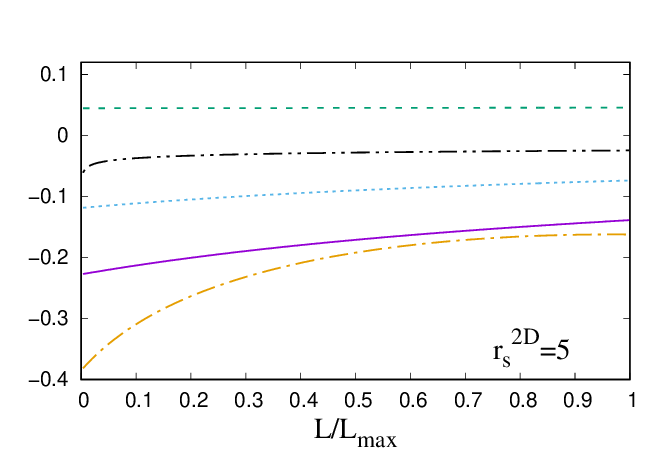}
\caption{The quasi-2D IBM ingredients for the ACII functionals:
the exact exchange per particle $E_x(L)/N$, $W_\infty(L)/N$ of Eq. \eqref{eql3},
$W'_\infty(L)/N$ of Eq. \eqref{eql5}, and the $W'_0(L)/N$ for the 2D bulk parameters
$r_s^{2D}=1$ (upper panel), $r_s^{2D}=3$ (middle panel), and $r_s^{2D}=5$ (lower panel),
respectively.
}
\label{f3}
\end{figure}
%
%
\begin{figure}[!hbt]
\includegraphics[width=\columnwidth]{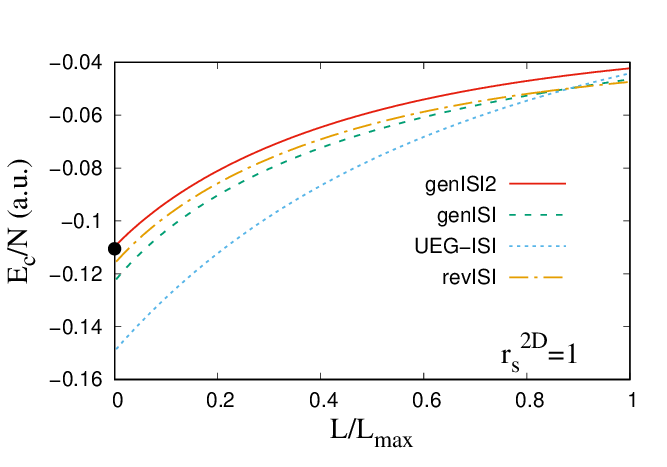}
\includegraphics[width=\columnwidth]{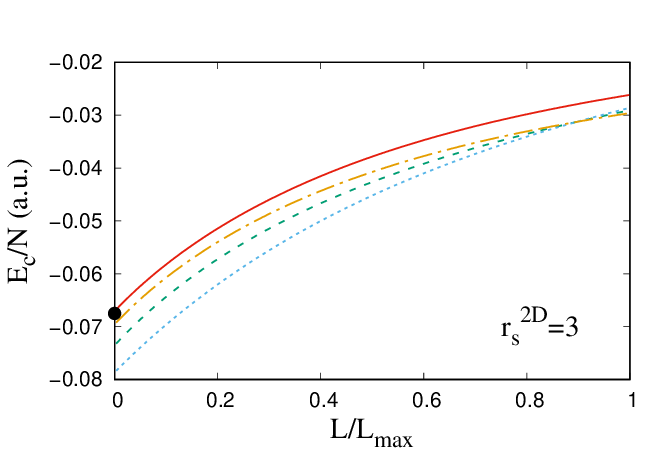}
\includegraphics[width=\columnwidth]{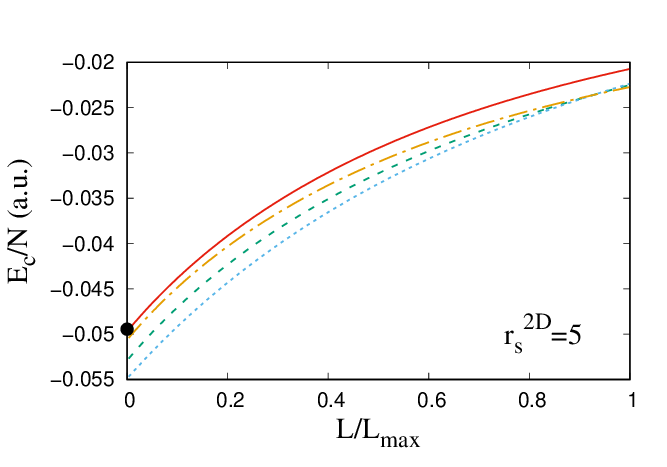}
\caption{Correlation energy per particle ($\epsilon_c$) of
the IBM quasi-2D electron gas of fixed 2D electron density ($r_s^{2D}=1$ upper panel, $r_s^{2D}=3$ (middle
panel), and $r_s^{2D}=5$ (lower panel), respectively),
as a function of the normalized quantum-well thickness $L/L_{max}$. The dots at $L=0$ represent the exact
correlation energies of the 2D UEG.
}
\label{f4}
\end{figure}
%
%

In Fig. \ref{f3}, we show all these ingredients for $r_s^{2D}=1$, 3 and 5, respectively. At high-densities (for 
$r_s^{2D}=1$), we observe that $W_\infty/N \le E_x/N \le W'_0/N \le 0 \le W'_\infty/N$, while at 
low-densities (for $r_s^{2D}=5$), the pattern is different $W'_0/N \le W_\infty/N \le E_x/N \le 0 \le W'_\infty/N$. In the figure, we also show $\epsilon_c^{LDA}$, for a better comparison with the other ingredients. Note that, $\epsilon_c^{LDA}\propto \ln(r_s)$ has a logarithmic divergence at $L\rightarrow 0$, where the 3D bulk parameter is going to vanish ($r_s\rightarrow 0$) due to the 3D high-density regime.  
   
Next, in Fig. \ref{f4}, we show the quasi-2D IBM correlation energy per particle from several ACII functionals in the whole quasi-2D regime (i.e., $0\le L/L_{max}\le 1$), for $r_s^{2D}=1$, 3 and 5, respectively. We compare the genISI, genISI2 and UEG-ISI with the revISI functional, which we have found to be the actual state-of-the-art functional for predicting the 2D correlation energy at $L\rightarrow 0$. We observe that genISI2 performs even better than revISI, being significantly better than genISI. On the other hand, the UEG-ISI underestimates the 2D correlation energy, being accurate only in a mild quasi-2D regime (i.e. $L/L_{max}\ge 0.6$). Nevertheless, we recall that the quasi-2D IBM is a very difficult test for any XC functional, and for example, the random phase approximation (RPA) performs worse than UEG-ISI (see Fig. 3 of Ref. \onlinecite{constantin2016simple}). On the other hand, the semilocal correlation functional, e.g.  PBE \cite{perdew1996generalized},  
can not describe the moderate and strong quasi-2D IBM regimes, being accurate at $L/L_{max}=1$ but approaching 0 at $L/L_{max}=0$ \cite{constantin2019correlation}. 
\subsection{Two-dimensional uniform electron gas (2D UEG):}
\label{sec35l}

We test the ACII functionals for the 2D UEG, where all the ingredients are known and discussed above; 
$W_0^{2D}/N=-4\frac{\sqrt{2}}{3\pi}\frac{1}{r_s^{2D}}$, $E_c^{GL2-2D}/N=-0.1925$ Ha,
$W_\infty^{2D}/N=\Big( \frac{8}{3\pi}-2 \Big)\frac{1}{r_s^{2D}}$, and 
$W^{'2D}_\infty/N=\frac{1}{2 (r_s^{2D})^{3/2}}$.

\begin{figure}[!hbt]
\includegraphics[width=\columnwidth]{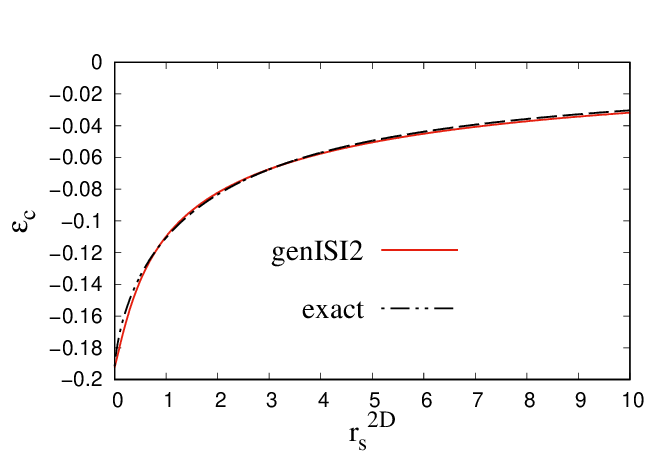}
\includegraphics[width=\columnwidth]{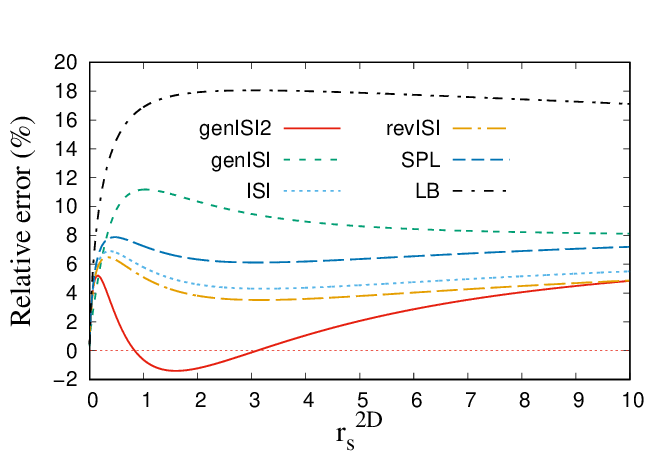}
\caption{Upper panel: 2D UEG correlation energy per particle $\epsilon_c$ versus the 2D bulk parameter $r_s^{2D}$. The exact curve is the 2D LDA correlation energy parametrization of Ref. \onlinecite{attaccalite2002correlation}.
\newline
Lower panel: Relative errors (in \%) of the correlation energy of the 2D UEG ($100\times(approx-exact)/exact$) obtained from various ACII functionals. }
\label{f5}
\end{figure}
%

In the upper panel of Fig. \ref{f5}, we show that genISI2 is remarkably close to the 2D exact LDA correlation energy per particle \cite{attaccalite2002correlation} for any value of the 2D bulk parameter $r_s^{2D}$, even in the high-density limit. In the lower panel of Fig. \ref{f5}, we report the relative errors (in \%) of the ACII functionals. The best performance is found for genISI2, followed by revISI and ISI, while LB and genISI give the worst results. 

To better quantify these results, let us consider the following integrated MARE (iMARE): 
\begin{equation}
\label{eqerr123}
\text{iMARE}=\frac{1}{(10-b)}\int^{10}_b dr_s 
\frac{|\epsilon_c^{approx}(r_s)-\epsilon_c^{exact}(r_s)|}{|\epsilon_c^{exact}(r_s)|} \times 100,
\end{equation}
where we take (b=0) for the 2D-UEG. We also compute Eq. \eqref{eqerr123} for the 3D-UEG with the choices (b=0) and (b=1). We recall that the considered ACII functionals are not accurate for the 3D-UEG in the limit $r_s\rightarrow 0$, where $\epsilon_c^{exact}\rightarrow 0.031091 \ln(r_s)-0.0469203$, while $\epsilon_c^{ISI},\epsilon_c^{revISI}\sim 1/\sqrt{r_s}$, $\epsilon_c^{UEG-ISI}\rightarrow -0.086$. In contrast, SPL and LB functionals perform as $\rightarrow 1/r_s$ not being integrable in the high-density limit. Nevertheless, we note that most bulk metals have $1\le r_s \le 10$. 
\begin{table}
\caption{\label{tabUEG} The iMARE of Eq. \eqref{eqerr123} for the 
2D-UEG and 3D-UEG. Best result is shown in bold style.}
\begin{tabular}{cccc}
\hline
&  2D-UEG  &   \multicolumn{2}{c}{3D-UEG} \\ 
\cmidrule(lr){2-2} \cmidrule(lr){3-4}
 &   b=0       & b=1 & b=0 \\ \hline
ISI & 5.0 & 45.0 & 59.4 \\
revISI & 4.3 & 27.3 & 37.7 \\
SPL & 6.6 & 241.5 & $\infty$ \\
LB & 17.2 & 241.5 & $\infty$ \\
UEG-ISI & 18.9 & \bf{0.9} & \bf{2.4}\\
genISI & 8.9 & \bf{0.9} & \bf{2.4} \\
genISI2 & \bf{2.5} & \bf{0.9} & \bf{2.4} \\
\hline
\end{tabular}
\end{table}
The iMARE values of Eq. \eqref{eqerr123} for both the 2D- and 3D- UEG are reported in Table \ref{tabUEG} from the considered ACII functionals. 

Thus, we conclude this subsection by noting that the improvement of genISI2 over the genISI is substantial for both quasi-2D IBM and 2D UEG and, corroborated with its 3D UEG accurate behavior, can make this functional attractive for various solid-state applications.   

\subsection{Finite systems}
\label{SEC:finite}
\subsubsection{Role of Reference orbitals}
The genISI2 functional has been parametrized on exact ingredients, reported in Tab. \ref{ta1}.
For practical applications, however, such exact values are not known. 
In  Tab. \ref{tab:reforb}, we report accurate results for the Harmonium ($\kappa=1/4$), He, and Ne atoms as obtained from accurate FCI (or CCSD(T)) density (as described in section \ref{seccd3}). 
Our results for $W_0$ and $E_c^{GL2}$ for the Harmonium and He atom reproduce previous exact 
results from the literature obtained in different ways, see Tab. \ref{ta1}. 
Data for $W_\infty$ and $W'_\infty$ in Tab.  \ref{tab:reforb} have been computed with the hPC functional with the accurate density.

The second row (labelled with $\partial_i$) of each section of Tab. \ref{tab:reforb} reports the relative derivatives of $E_c^{genISI2}$ with respect to the various ingredients, i.e.
\begin{equation}
\label{eq:parti}
\partial_i=\frac{1}{W_i} \frac{\partial E_c^{genISI2}}{\partial W_i}
\end{equation}
where $W_i$ is one of the four ingredients ($W_0$, $W_\infty$, $W'_\infty$ and $E_c^{GL2}$)
The $E_c^{GL2}$ ingredient has the largest coefficients, particularly for atomic systems.
For those systems, thus the $W'_0=2E_c^{GL2}$ ingredient has a very large impact on the final correlation energy. For this reason, its accurate calculation is much more important than the SCE ingredient.


Then, we compute the ACII ingredients consider different input orbitals and densities (S-VWN \cite{vwn_ueg}, PBE, HF, and OEPx), which are routinely available.
We can see that the type of input orbitals have a drastic effect on the $E_c^{GL2}$ energy (relative deviations -38\%$\dots$+11\%). In contrast, the other quantities are not much 
affected (with a relative deviation of few percent, and often much smaller).  
In fact, the value
of the $E_c^{GL2}$ energy is directly related to the HOMO-LUMO energy gap (last column of Table \ref{tab:reforb}),
whereas the other ingredients depend only on the ground-state density (matrix), which does not change much. HF largely underestimates $E_c^{GL2}$, whereas PBE overestimates it.
The best agreement with the reference $E_c^{GL2}$ is obtained using OEPx orbital, and in fact, OEPx gives an energy gap closer to the reference one \cite{smiga2022selfconsistent,tls2}. More sophisticated approaches \cite{grabowski2014orbital,smiga2022selfconsistent} are more computationally demanding or numerically cumbersome.

Thus, the calculation of genISI2 correlation energies on OEPx orbitals seems the best choice, considering both the accuracy and the computational cost. Then, for all finite systems, the considered ACII functionals have been evaluated using OEPx orbitals.

If the genISI2 functional is used with very different orbitals (PBE, HF), a proper scaling of the $E_c^{GL2}$ must be used. Otherwise, the results can be quite nonphysical.
\begin{table*}[hbt]
\caption{\label{tab:reforb} The ACII ingredients for different input densities, in case of the the Harmonium ($\kappa=1/4$), Helium, and Neon atoms. We used the uncontracted aug-cc-pV6Z basis set, except for Harmonium, where a special basis set is utilized \cite{B926389F}.
Values of  $W_\infty$ and $W'_\infty$ are computed with the hPC functional \cite{smiga2022selfconsistent}. The last two columns report the genISI2 correlation energy and the HOMO-LUMO gap. We also report the relative derivatives ($\partial_i$, see Eq. \eqref{eq:parti}) for FCI / CCSD(T) densities.}
\begin{tabular}{c|ccccrr}
\hline
      & $W_0$ & $W_\infty$  & $W'_\infty$ & $E_c^{GL2}$ & $E_c^{genISI2}$  & Gap (eV)\\
 \hline
   \multicolumn{7}{c}{Harmonium} \\
   \hline
     FCI & -0.515 & -0.743 & 0.207 & -0.0496 & -0.0370 &11.47\\ 
  $\partial$    &    -15.7\%    &  15.7\%      &  8.4\%    & 18.9\%     &         & \\
  \hline
  S-VWN  & -0.511 (-0.8\%)& -0.738 (-0.7\%)& 0.204(-1.4\%) & -0.0497(+0.2\%) & -0.0371 &11.32\\ 
  PBE   & -0.513 (-0.4\%) & -0.741 (-0.3\%)& 0.205(-1.0\%) & -0.0498(+0.4\%) & -0.0372 &11.42\\
  HF  & -0.515 (0\%)     & -0.743 (0\%)&  0.208(0.5\%) & -0.0304(-38\%) & -0.0283 &20.84\\
  OEPx & -0.515 (0\%)      & -0.743 (0\%)& 0.208(0.5\%) & -0.0494(-0.4\%) & -0.0368 &11.45\\

 \hline
     \multicolumn{7}{c}{Helium} \\
   \hline  
       FCI & -1.024 & -1.491 & 0.644 & -0.0480 & -0.0414 & 20.40\\ 
       $\partial$ & -6.9\%  & +6.9\%  & +2.6\% & +27.1\% \\
  \hline     
  S-VWN  & -0.998 (-2.5\%)       & -1.454 (-2.5\%)   & 0.619 (-3.9\%)    & -0.0533 (+11\%) & -0.0440 & 16.90 \\ 
  PBE   & -1.013 (-1.1\%)        & -1.476 (-1.0\%)   & 0.634  (-1.6\%)   & -0.0514 (+7.1\%)      & -0.0431       & 17.00 \\
  HF    & -1.026 (+0.2\%)         & -1.492 (+0.1\%)  & 0.645 (+0.2\%)    & -0.0366  (-23\%)     & -0.0345       & 27.46 \\
  OEPx  & -1.026 (+0.2\%)          & -1.492 (+0.1\%)   & 0.646 (+0.3\%) & -0.0478 (-0.4\%) & -0.0412 & 20.77 \\

 \hline
    \multicolumn{7}{c}{Neon} \\
    \hline
    CCSD(T)   & -12.078      &  -20.051     &   23.041    &  -0.4741   &   -0.3884 &     17.12 \\
    $\partial$ &    -2.6\%          &    +2.5\%          &   +0.3\%          &   +29.5\%        &           &           \\ 
    \hline
 S-VWN & -12.008 (-0.6\%)  &  -19.964 (-0.4\%) & 22.952 (-0.4\%) & -0.499 (+5.3\%) &  -0.403  & 15.53 \\
 PBE   & -12.044 (-0.3\%)  & -20.011 (-0.2\%)  & 23.016 (-0.1\%) & -0.491 (+3.6\%) & 	-0.421    &  15.15 \\
 HF    & -12.108 (+0.2\%) & 	-20.076 (+0.1\%)  & 23.045 (0\%)    & -0.367 (-22.6\%)  & 	-0.320    & 27.39 \\
 OEPx  &  -12.104 (+0.2\%) &  -20.078 (+0.1\%)  & 23.044 (0\%)    & -0.4631(-2.3\%) & -0.3819 &  18.44 \\ 
 \hline
\end{tabular}
\end{table*}

\subsubsection{Harmonium}
\label{ddhook}

\begin{figure}[!hbt]
\includegraphics[width=\columnwidth]{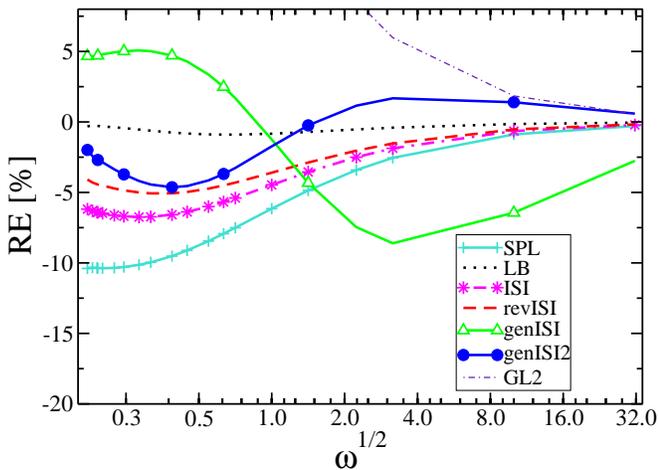}
\caption{Relative error on correlation energies of Harmonium at various values of $\omega$ computed with OEPx orbitals for several ACII functionals using the hPC model for the strong-interaction functionals. The errors have been computed with respect to FCI data obtained in the same basis set \cite{B926389F}.}
\label{fig:hook}
\end{figure}

 In Fig. \ref{fig:hook}, we show the relative errors (in \%) given by ACII functionals for Harmonium within a broad interval of frequencies $0.03\le \omega \le 1000$.
Calculations are done using OEPx orbitals, which gives quite an accurate density in this case \cite{smiga2020modified,smiga2022selfconsistent}.

In the strongly-correlated regime (i.e., $\omega\le 0.5$), the best result is given by the LB followed by the genISI2
(with $-2\% \lessapprox RE \lessapprox -0.5\%$), while the worst is SPL (with the error below 10 \%). GL2 is very inaccurate in this region.
On the other hand, for the tighter bound electrons ($\omega\ge 1$) until the high-density limit ($\omega\gtrapprox 100$), the genISI2 gives a similar 
performance to other methods being much better than genISI, which is the worst.

Overall, the genISI2 (MARE = 2.77\%) improves over the genISI (MARE = 4.55\%), being better than other functionals, such as SPL (MARE = 7.66\%), ISI (MARE = 5.09\%)
 and revISI (MARE = 3.79\%). The LB gives the smallest MARE = 0.52\% which is consistent with other predictions \cite{kooi2018local}.

\subsubsection{H$_2$ Dissociation}
\label{sec:h2diss}

\begin{figure}
\includegraphics[width=\columnwidth]{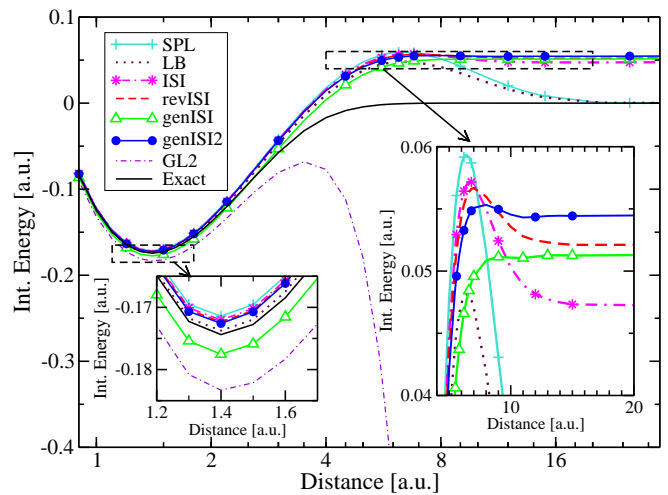}
\caption{ $H_2$ dissociation curve computed in a spin-restricted formalism with several ACII functionals, using OEPx orbitals. The exact curve is also reported.
}
\label{fh2}
\end{figure}

In Fig. \ref{fh2}, we report the $H_2$ dissociation curve using a spin-restricted formalism
for ISI, revISI, genISI, and genISI2, as well as 
the FCI
(exact) results.
We used OEPx orbitals and aug-cc-pV5Z basis-set.
Note that, for two-electron systems, OEPx calculation can be done exactly, 
as the exchange potential is half of the Hartree potential \cite{weimer}.
In addition, the singly excited term vanishes \cite{grabowski2014orbital}. Fig. \ref{fh2} highlights three important regions:
\begin{itemize}
\item[i)] at equilibrium distance (see inset on the left), all methods are quite accurate. The highest accuracy is obtained from LB, genISI2, ISI, and revISI, while 
genISI over-binds and SPL under-binds.
\item[ii)] at the dissociation limit, all ISI methods converge to a constant D, in contrast to GL2, which diverges.
The constant D can be directly computed as
\begin{eqnarray}
D&=&E[H_2]-2E[H] = 2(E[H_{1/2,1/2}]-E[H]), 
\label{eqdiss}
\end{eqnarray}
where $E[H_{1/2,1/2}]$ is the total energy of the strongly correlated hydrogen atom \cite{cohen08} with half electron spin-up spin-down occupation.
The resulting energies are reported in Tab. \ref{tab:h2ene}.
In the limit $E_c^{GL2}=-\infty$, genISI and genISI2 reduce to the same
UEG-ISI functional, and thus the same energy (-0.47239 Ha) is obtained. An almost perfect agreement is obtained with the SPL and LB functionals. We recall that SPL and LB perform identical (as $W_\infty$), in the limit $E_c^{GL2}=-\infty$.
However, it should be considered that this is just
an error cancellation between the approximated electronic density
and the (approximated) hPC SCE model.
Using the exact density, in fact, the SPL and LB give a too low energy.
The difference between SPL (LB) and the other ACII functionals originates from the $W'_{\infty}$ contribution which does not vanish using approximated expressions, such as PC or hPC. The exact  $W'_{\infty}$ is zero for $E[H_{1/2,1/2}]$, and, in this case, all the ACII functionals will be identical to SPL (LB).


\item[iii)] in the region around 4-8 a.u. (see right inset), a  large repulsive bump emerges in the case of SPL and LB, and also for ISI and revISI. On the other hand, genISI and genISI2 show a much smoother curve. The bump is related to deficiencies of XC energy expression to describe fully the regions where static and dynamic correlation effects interplay. This issue has already been discussed in the literature in different contexts \cite{fuchs2005desc,BGE2,vuckovic2016XC,vuckovic2017simple}. We note that proper dissociation for H$_2$ can be recovered using highly nonlocal \cite{vuckovic2017simple} forms of AC or by using simple AC formulas \cite{peach2008adiabatic} with accurate input ingredients for $W_\infty$ and  $W'_\infty$.
\end{itemize}

In summary, the genISI2 provides a quite accurate description of the H$_2$ dissociation curve, performing significantly better than genISI near the equilibrium distance, while at large distances between the H atoms, both genISI2 and genISI are similar, recovering the UEG-ISI functional. Their behaviors strongly depend on the quality of the approximations for $W_\infty$ and $W'_\infty$. However, they yield the exact result when the exact SCE $W_\infty$ and $W'_\infty$ ingredients are used.     


\begin{table}
\caption{\label{tab:h2ene} Total energy of $H_{1/2,1/2}$ and the hydrogen atom (H) for different methods. Results for $H_{1/2,1/2}$ are reported with both OEPx and exact orbitals. }
        \begin{tabular}{ccccc}
        \hline
       &  \multicolumn{2}{c}{$H_{1/2,1/2}$}  &    H  & MARE\%\\ 
       \cmidrule(lr){2-3} \cmidrule(lr){4-4}
          &   @OEPx     & @Exact     &   @Exact        &   \\ \hline
   OEPx    & -0.35771   &   -0.34374  & -0.5     & 28.48 \%  \\
   ISI     & -0.47591    & -0.48120  & -0.5       & 4.82   \%  \\
   revISI  & -0.47350   &  -0.47738 & -0.5        & 5.30   \% \\
   genISI  & -0.47239   &  -0.47536  &  -0.49840  &  5.20 \% \\
   genISI2 & -0.47239   & -0.47536 & -0.5          &  5.52 \% \\
   SPL     &  -0.49994  & -0.51676 &  -0.5        & 0.01   \% \\
   LB      &  -0.49994  &  -0.51676 &  -0.5       & 0.01   \%  \\
   \hline
   Exact  &  \multicolumn{2}{c}{-0.5}       &   -0.5 \\
   \hline
  
   \hline
\hline
        \end{tabular}
\end{table}
%
\subsubsection{Molecular systems}
\label{ddmol}

First, we show in Table \ref{taTE} the total energies (in Ha) for a test-set of  5 atoms and 11 small molecules, previously used in \textit{ab initio} DFT calculations \cite{grabowski2014orbital}.
Although total energies are not very important in practical chemical applications, they are essential observables and are especially useful as indicators of the quality of the ACII interpolation functions.

\begin{table*}[htbp]
\caption{Total energies (Ha) for several ACII functionals computed using OEPx orbitals, w.r.t. CCSD(T) reference data. The last two rows report the MAE (in mHa) and MARE errors. For comparison, we also report the MP2 and PBE total energies. The best results from ACII functionals are boldfaced. 
}
\begin{tabular}{crrrrrrrrr}
\hline \hline
System & CCSD(T) & MP2 & PBE & LB & SPL & ISI & revISI & genISI & genISI2 \\  \hline
He & -2.9025 & -2.8970 & -2.8929 & -2.9032 & -2.9015 & -2.9019 & -2.9021 & {\bf -2.9024} & -2.9028 \\ 
Be & -14.6623 & -14.6424 & -14.6299 & -14.6767 & -14.6728 & -14.6710 & -14.6702 & -14.6745 & {\bf -14.6655} \\ 
Ne & -128.9000 & -128.8924 & -128.8659 & -128.9465 & -128.9363 & -128.9273 & -128.9231 & -128.9287 & {\bf -128.9083} \\ 
Mg & -199.8282 & -199.8157 & -199.9539 & -199.8690 & -199.8657 & -199.8656 & -199.8656 & {\bf -199.8637} & -199.8663 \\ 
Ar & -527.4575 & -527.4336 & -527.3451 & -527.5440 & -527.5309 & -527.5148 & -527.5071 & -527.5089 & {\bf -527.4882} \\ 
HF & -100.3958 & -100.3861 & -100.3859 & -100.4534 & -100.4419 & -100.4315 & -100.4266 & -100.4323 & {\bf -100.4096} \\ 
CO & -113.2574 & -113.2339 & -113.2323 & -113.3680 & -113.3476 & -113.3277 & -113.3185 & -113.3260 & {\bf -113.2872} \\ 
H$_2$O & -76.3869 & -76.3720 & -76.3754 & -76.4457 & -76.4327 & -76.4208 & -76.4152 & -76.4211     & {\bf -76.3958} \\ 
H$_2$ & -1.1727 & -1.1650 & { -1.1662} & {\bf -1.1717} & -1.1695 & -1.1702 & -1.1705 & -1.1720 & -1.1711 \\ 
He$_2$ & -5.8051 & -5.7939 & { -5.7859} & {\bf -5.8040} & -5.8008 & -5.8017 & -5.8021 &-5.8018 & -5.8038 \\ 
Cl$_2$ & -919.7703 & -919.7222 & -920.0431 & -919.9427 & -919.9302 & -919.9241 & -919.9211 &-919.9300 & {\bf -919.9098} \\ 
N$_2$ & -109.4763 & -109.4562 & -109.4507 & -109.6093 & -109.5861 & -109.5611 & -109.5497 &-109.5548 & {\bf -109.5134} \\ 
Ne$_2$ & -257.8000 & -257.7850 & -257.7319 & -257.8931 & -257.8727 & -257.8548 & -257.8463 &-257.8574 & {\bf -257.8168} \\ 
HCl & -460.5093 & -460.4823 & -460.6385 & -460.5920 & -460.5855 & -460.5823 & -460.5807 &-460.5854 & {\bf -460.5746} \\ 
NH$_3$ & -56.5233 & -56.5016 & -56.5072 & -56.5715 & -56.5588 & -56.5486 & -56.5439 &-56.5516 & {\bf -56.5253} \\
C$_2$H$_6$ & -79.7641 & -79.7145 & -79.7270 & -79.8343 & -79.8124 & -79.7988 & -79.7923 &-79.8106 & {\bf -79.7630} \\ 
 &  &  &  &  &  &  &  &  \\
MAE &  & 19.87 & 58.46 & 63.60 & 53.16 & 43.96 & 41.76 &44.86 & \textbf{24.86} \\ 
MARE &  & 0.08\% & 0.11\% & 0.06\% & 0.06\% & 0.05\% & 0.041\% & 0.04\% & \textbf{0.02\%} \\ \hline \hline
\end{tabular}
\label{taTE}
\end{table*}

The genISI2 (with MAE $\approx$ 25 mHa and MARE = 0.02 \%) gives the best performance among ACII functionals, being almost twice better than revISI (MAE $\approx$ 42 mHa and MARE = 0.04 \%), which is slightly more accurate than ISI and genISI. The MP2 method yields here MAE and MARE of about 19.87 mHa and 0.09\%, respectively, which indicates the large improvement of genISI2 functional with respect to its counterparts. The worst performance is given by LB
(MAE $\approx$ 64 mHa and MARE = 0.06 \%) and SPL (MAE $\approx$ 53 mHa and MARE = 0.06 \%) functionals, which yield a result similar to the PBE functional (MAE $\approx$ 58 mHa and MARE = 0.11 \%).

We note that the total energy test is important from the theoretical point of view because it can also show if a given functional that is accurate for atomization energies of molecules relies on an error cancellation (e.g., most semilocal GGA and meta-GGA XC functionals).  

\begin{figure}
\includegraphics[width=\columnwidth]{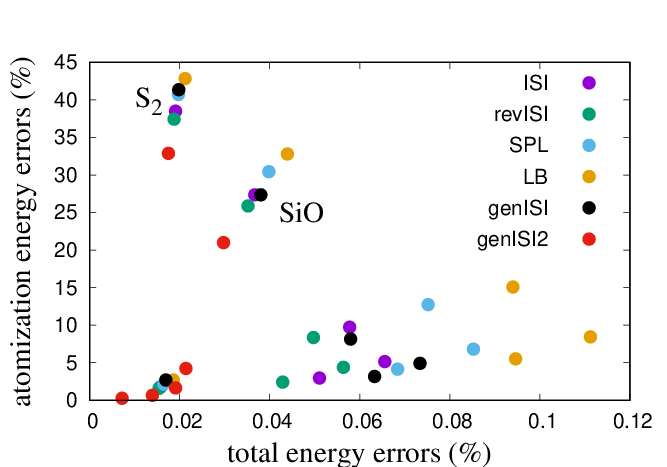}
\caption{Absolute relative errors (ARE) on the total energies versus the ARE on atomization energies for the molecules of the AE6 test (SiH$_4$, SiO, S$_2$, C$_3$H$_4$, C$_2$H$_2$O$_2$ and C$_4$H$_8$). We use the CCSD(T) as reference data (see also Table \ref{taAE6}).
}
\label{fAE6}
\end{figure}
\begin{table*}[htbp]
\caption{Atomization energies (in kcal/mol) for the AE6 test,
computed using OEPx orbitals. The last two rows report
the error statistics (MAE and MARE) computed w.r.t. CCSD(T) reference data. The 
best results are highlighted in bold.}
\begin{tabular}{crrrrrrrr}
\hline \hline
 & CCSD(T) & LB & SPL & ISI & revISI & genISI & genISI2 \\  \hline
SiH$_4$ & 318.8 & 327.4 & 325.7 & 324.5 & 323.9 & 327.4 & \bf{320.9} \\ 
SiO & 182.0 & 241.7 & 237.4 & 231.8 & 229.1 & 231.8 & \bf{220.2} \\ 
S$_2$ & 94.3 & 134.7 & 132.7 & 130.6 & 129.6 & 133.3 & \bf{125.3} \\ 
C$_3$H$_4$ & 690.8 & 749.1 & 737.8 & 726.4 & 721.2 & 724.8 & \bf{702.2} \\ 
C$_2$H$_2$O$_2$ & 618.1 & 711.4 & 696.9 & 678.3 & 669.8 & 668.5 & \bf{644.3} \\ 
C$_4$H$_8$ & 1130.5 & 1192.9 & 1177.3 & 1164.1 & 1158.0 & 1166.4 & \bf{1133.6} \\ 
 &  &  &  &  &  &  \\ 
MAE &  & 53.8 & 45.6 & 36.9 & 32.9 & 36.3 & \textbf{18.7} \\ 
MARE &  & 17.9\% & 16.2\% & 14.3\% & 13.4\%&14.6\% & \textbf{10.1\%} \\ 
\hline \hline
\end{tabular}
\label{taAE6}
\end{table*}
Next, let us consider the popular AE6 benchmark that is representative for the atomization energies of a large molecular database \cite{lynch2003robust}.
Note that we consider the error vs CCSD(T) results in the same basis-set and 
not versus experimental results.
In this way, a more simple and direct comparison can be performed without
considering complete-basis-set extrapolation issues.
In Table \ref{taAE6}, we show the AE6 atomization energies (in kcal/mol) from the ACII functionals. The genISI2 improves over all the other considered functionals, reducing the MAE from $\gtrapprox 33$ kcal/mol to $\approx 19$ kcal/mol.
This is a quite significant improvement. However, we should point out
that other conventional approaches give much lower errors.
For example, MP2 yields MAE=12.1 and MARE=3.71\%, see Ref. \onlinecite{smiga2020modified}.
In fact, the calculation of atomization energies with ACII functional
based on KS orbitals is still a challenge, such that the genISI2 functional
is an important step in this direction.

To better visualize the performance of each functional, we report in Fig. \ref{fAE6} the absolute relative errors (ARE) on total energies versus the ones on atomization energies for the six molecules of the AE6 test. As shown, the genISI2 gives a systematic improvement for all the molecules. Thus, the maximum error on atomization energies is for the S$_2$ dimer (ARE=32.9 \%), followed by the SiO molecule (ARE=21\%), while all the remaining molecules have ARE below 4\%. On the other hand, the total energies are remarkably accurate, the worst case being the SiO molecule with ARE=0.0298 \%.

In Table. \ref{taIP32}, we show the results for the ionization potentials
of the 32 systems (atoms and small molecules) reported in Ref. \onlinecite{SmigaIP}.
The best result is found from genISI2 (MARE=3.5 \%, MAE=0.44 eV), which gives a systematic improvement over the other functionals, providing the best result for 27 systems out of 32. After that, ISI, revISI and genISI give almost similar performances (with $4.6 \% \le \rm{MARE} \le 5 \%$ and $0.57\; \rm{eV} \le \rm{MAE} \le 0.61\; \rm{eV}$), while SPL and LB have the worst performances, with MAE$\ge 0.7$ eV. 

\begin{table}[htbp]
\caption{Ionization potentials  (in eV) computed from an energy difference obtained for several ACII expressions.
The last two rows report
the error statistics (MAE (eV) and MARE (\%)) computed w.r.t. reference data. The smallest error of each row is highlighted in bold.}
\begin{tabular}{lccccccc}
\hline \hline
	&	Ref.	&	LB	&	SPL	&	ISI	&	revISI	&	genISI	&	genISI2 \\ \hline
Ar	&	-15.63	&	-16.15	&	-16.10	&	-16.01	&	-15.97	&	-15.94	&	\bf{-15.89} \\
Be	&	-9.31	&	-9.72	&	-9.63	&	-9.58	&	-9.55	&	-9.76	&	\bf{-9.38} \\
C$_2$H$_2$	&	-11.43	&	-12.36	&	-12.21	&	-12.01	&	-11.91	&	-11.88	&	\bf{-11.64} \\
C$_2$H$_4$	&	-10.63	&	-11.40	&	-11.25	&	-11.07	&	-10.99	&	-11.00	&	\bf{-10.72} \\
C$_2$H$_6$	&	-13.01	&	-12.70	&	-12.69	&	-12.70	&	-12.70	&	\bf{-12.74}	&	-12.71 \\
CH$_2$CF$_2$	&	-10.61	&	-11.00	&	-10.93	&	-10.84	&	-10.79	&	-10.78	&	\bf{-10.67} \\
CH$_3$CN	&	-10.75	&	-11.94	&	-11.80	&	-11.61	&	-11.52	&	-11.47	&	\bf{-11.28} \\
CH$_4$	&	-14.37	&	-14.65	&	-14.58	&	-14.51	&	-14.49	&	-14.52	&	\bf{-14.37} \\
CHF$_3$	&	-14.59	&	-15.75	&	-15.71	&	-15.67	&	-15.65	&	-15.67	&	\bf{-15.60} \\
Cl$_2$	&	-11.45	&	-11.05	&	-11.06	&	-11.07	&	-11.08	&	-11.07	&	\bf{-11.10} \\
CO$_2$	&	-13.70	&	-14.93	&	-14.80	&	-14.62	&	-14.54	&	-14.49	&	\bf{-14.33} \\
CO	&	-13.94	&	-14.62	&	-14.54	&	-14.46	&	-14.42	&	-14.42	&	\bf{-14.31} \\
CS	&	-11.27	&	-14.16	&	-14.02	&	-13.82	&	-13.72	&	-13.73	&	\bf{-13.45} \\
FCN	&	-13.65	&	-15.83	&	-15.68	&	-15.46	&	-15.36	&	-15.29	&	\bf{-15.11} \\
H$_2$CO	&	-10.83	&	-11.70	&	-11.58	&	-11.42	&	-11.35	&	-11.32	&	\bf{-11.16} \\
H$_2$CS	&	-9.29	&	-10.47	&	-10.39	&	-10.28	&	-10.22	&	-10.22	&	\bf{-10.07} \\
H$_2$O	&	-12.50	&	-13.19	&	-13.08	&	-12.92	&	-12.85	&	-12.83	&	\bf{-12.66} \\
HCCF	&	-11.50	&	-12.08	&	-11.97	&	-11.82	&	-11.75	&	-11.71	&	\bf{-11.57} \\
HCl	&	-12.59	&	-13.15	&	-13.11	&	-13.07	&	-13.06	&	-13.10	&	\bf{-12.98} \\
HCN	&	-13.90	&	-15.74	&	-15.54	&	-15.26	&	-15.14	&	-15.06	&	\bf{-14.80} \\
He	&	-24.48	&	-24.54	&	\bf{-24.50}	&	-24.51	&	-24.52	&	-24.51	&	-24.54 \\
He$_2$	&	-24.48	&	-24.62	&	-24.60	&	-24.58	&	-24.56	&	-25.17	&	\bf{-24.42} \\
HF	&	-15.96	&	-16.68	&	-16.58	&	-16.43	&	-16.37	&	-16.35	&	\bf{-16.18} \\
Mg	&	-7.57	&	-8.03	&	-8.00	&	-7.99	&	-7.98	&	-8.16	&	\bf{-7.92}  \\
N$_2$	&	-15.51	&	-13.96	&	-14.11	&	-14.52	&	-14.69	&	\bf{-15.00}	&	-14.95 \\
NCCN	&	-13.51	&	-13.34	&	\bf{-13.51}	&	-13.94	&	-14.12	&	-14.44	&	-14.43 \\
Ne$_2$	&	-21.34	&	-22.03	&	-21.94	&	-21.81	&	-21.75	&	-21.74	&	\bf{-21.59} \\
Ne	&	-21.47	&	-22.08	&	-21.98	&	-21.86	&	-21.80	&	-21.80	&	\bf{-21.64} \\
NH$_3$	&	-10.78	&	-11.31	&	-11.20	&	-11.08	&	-11.02	&	-11.02	&	\bf{-10.84} \\
OCS	&	-11.18	&	-12.30	&	-12.22	&	-12.09	&	-12.03	&	-11.99	&	\bf{-11.89} \\
P$_2$	&	-10.66	&	-10.38	&	-10.38	&	-10.38	&	-10.38	&	\bf{-10.38}	&	-10.37 \\
SiH$_4$	&	-12.78	&	-13.02	&	-12.99	&	-12.97	&	-12.96	&	-13.01	&	\bf{-12.91} \\ 
 &  &  &  &  &  &  \\ 
MAE & &	0.80	&0.71&	0.61&	0.57&	0.59&	\bf{0.44} \\
MARE(\%) & &	6.45 &	5.75&	4.95&	4.59&	4.74&	\bf{3.52}\\
\hline \hline
\end{tabular}
\label{taIP32}
\end{table}
%

\section{Summary and Conclusions}
\label{secconc}

\begin{figure}
\includegraphics[width=\columnwidth]{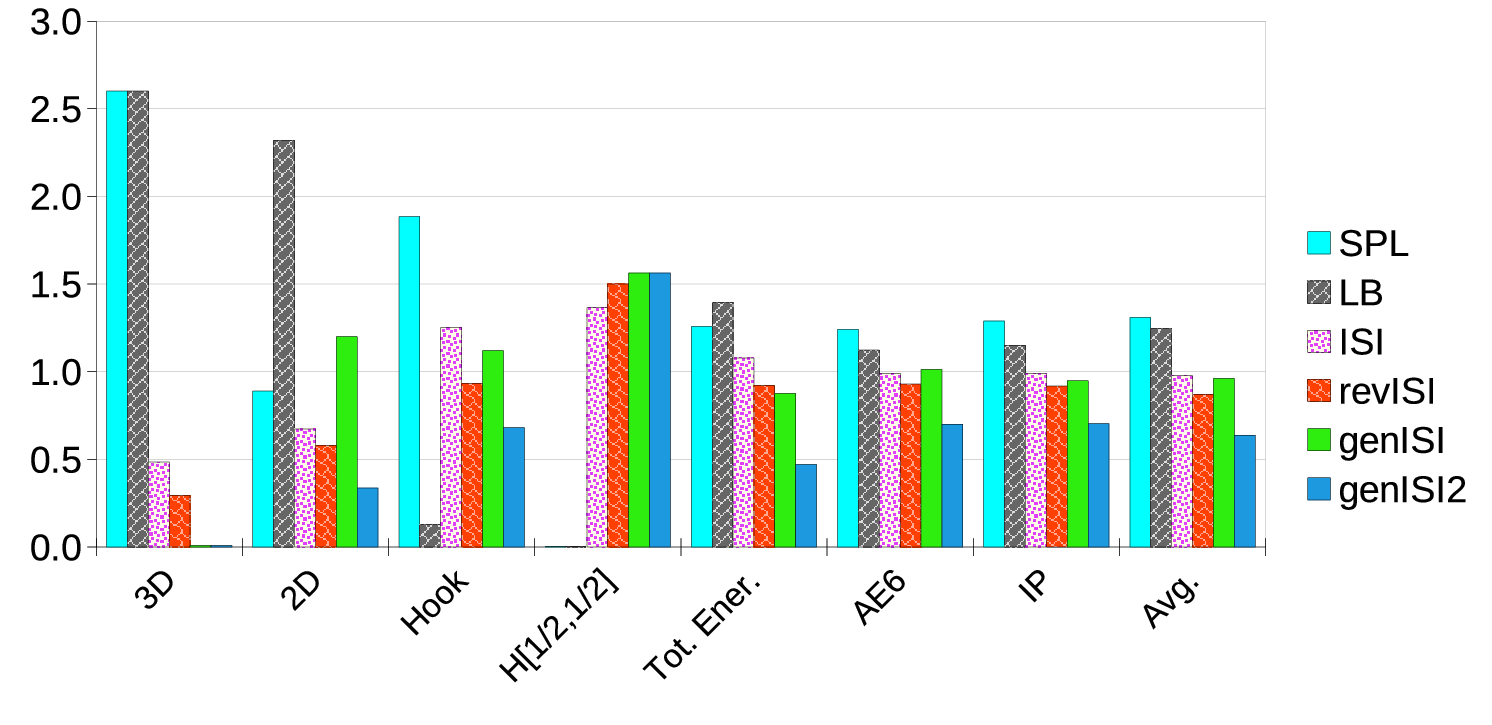}
\caption{
\label{fig:comp} Bar-plot of the normalized MARE for the 7 tests and the 6 ACII functionals considered in this work. The last section reports the normalized MARE averaged over all tests.}
\end{figure}

We have developed a novel ACII approach, denoted as genISI2, that satisfies the negativity constraint of the correlation energy and the exact conditions for the weak- and strong-interaction limits. When the GL2 correlation diverges, the genISI2 approach recovers the UEG-ISI functional, which is correct for the 3D uniform electron gas. On the other limit, i.e., when the GL2 correlation vanishes, the genISI2 correlation energy also correctly goes to zero, see Eq. \eqref{eq15t3d}.

The genISI2 functional is based on two parameters that have been fixed from the correlation energies of small systems, where the ingredients used for the weak- and strong-interaction limits ($W_0$, $W'_0$, $W^{SCE}_\infty$, and $W'^{SCE}_\infty$) are known (almost) exactly. Such a simple optimization procedure seems powerful and practical because the AC correlation integrand of the genISI2 functional $W_{\alpha,c}$ achieved good accuracy for both the Harmonium 
and Ne atoms, as reported in Fig. \ref{f1a}. Moreover, improved results are obtained for other very different systems, ranging from two-dimensional systems to atomization energies of molecules.

We compared genISI2 with all the currently available ACII functionals.
In Fig. \ref{fig:comp}, we report the normalized MARE 
\begin{equation}
\eta_i^j = \frac{\rm{MARE}_i^j}{\frac{1}{6}\sum_{j} \rm{MARE}_i^j},
\label{eq_err_bar}
\end{equation} 
where $MARE_i^j$ is the MARE of the $i$-th test for functional $j$.
The index $j$ runs over the 6 ACII functionals considered, and the index $i$
runs over the 7 tests considered, namely: 
3D UEG (see Tab. \ref{tabUEG}), 
2D UEG (see Tab. \ref{tabUEG}),
Harmonium (see Sect. \ref{ddhook}),
the strongly correlated hydrogen (see Tab. \ref{tab:h2ene}),
total energies of atoms and molecules (see Tab. \ref{taTE}),
atomization energies (see Tab. \ref{taAE6}), and
ionization potential (see Tab. \ref{taIP32}).

The first section of Fig. \ref{fig:comp} shows that for the 3D UEG, as also discussed in Ref. \onlinecite{constantin2023adiabatic}, SPL
and LB are very inaccurate, while revISI and ISI improve significantly.
Instead, by construction, genISI and genISI2 are almost exact.

For the quasi-2D 
IBM, which is one of the most severe tests for XC density functionals \cite{pollack2000evaluating},  
we have shown that the genISI2 performs remarkably well in the whole quasi-2D regime, see section \ref{sec43}. For the true 2D UEG (see the second section of Fig. \ref{fig:comp}), the genISI2 is the best functional. Such high accuracy for 3D, quasi-2D and 2D uniform electron gases should make it attractive for various electronic calculations of solid-state and condensed-matter physics. 

In the case of finite systems, we have shown that great care is required for the calculation of the GL2 correlation energy, being the most important ingredient to obtain the correct correlation energy, which depends strongly on the chosen ground-state orbital. While fully self-consistent calculations are, in principle, possible \cite{smiga2022selfconsistent}, a more computationally cheaper yet accurate approach is required to investigate large systems.
We have shown in Table \ref{tab:reforb} that Kohn-Sham OEPx orbitals can be a reliable choice to obtain accurate ingredients for the weak- and strong-interaction limits 
($W_0$, $W'_0$, $W_\infty$, and $W'_\infty$). Thus, all finite system calculations in this work have been performed using OEPx orbitals.
 
For the Harmonium (see third section of Fig. 
\ref{fig:comp}), genISI2 improves with respect to all ACII functional, except LB, which is almost exact. LB (and also SPL) is almost exact for the $H_2$ dissociation (see the fourth section of Fig. \ref{fig:comp}), whereas all the other ACII approaches give similar errors.
However, as discussed in section \ref{sec:h2diss}, for these strongly correlated electron systems, the accuracy of the results also strongly depends on the accuracy of $W_\infty$, and $W'_\infty$ (in this work approximated with hPC) and not only on the ACII interpolation formula. Using the exact SCE for $W_\infty$, and $W'_\infty$, all the ACII functionals will be exact.

For atoms and molecular systems, we have observed a systematic improvement of the genISI2 functional with respect to the other ACII approximations for total energies, atomization energies, and ionization potential, see last three sections of Fig. \ref{fig:comp}.
Those are hard tests, despite the fact that the final results are not yet competitive with the most accurate quantum-chemistry methods.
Note, in addition, that the genISI2 functional has been parametrized on exact values, whereas for the calculations
of real atoms and molecules, approximated quantities are used
(i.e., OEPx orbitals instead of exact ones, hPC model data instead of SCE).
Thus, further improvement of the results can be expected if this ACII functional is parametrized on those approximated ingredients.

Averaging over all tests, the genISI2 has $\eta=0.63$, a relevant improvement concerning ISI, revISI, and genISI, which are all above 0.85.
We thus expect that genISI2 can have broader applicability and find applications in different areas, ranging from real metal clusters to solid-state systems. 


\section*{Acknowledgments}
L.A.C. and F.D.S. acknowledge the financial support from
ICSC - Centro Nazionale di Ricerca in High Performance
Computing, Big Data and Quantum Computing, funded by 
European Union - NextGenerationEU - PNRR.
F.D.S acknowledges the financial support from
PRIN project no. 2022LM2K5X, Adiabatic Connection for Correlation in Crystals ($AC^3$) funded by
European Union - NextGenerationEU - PNRR.
S.\'S. acknowledges the financial support from the National Science Centre, Poland
(grant no. 2021/42/E/ST4/00096).

\appendix*
\section{The negativity condition of the correlation energy:}
\label{sec41A}
 %
\begin{figure}[hbt]
\includegraphics[width=\columnwidth]{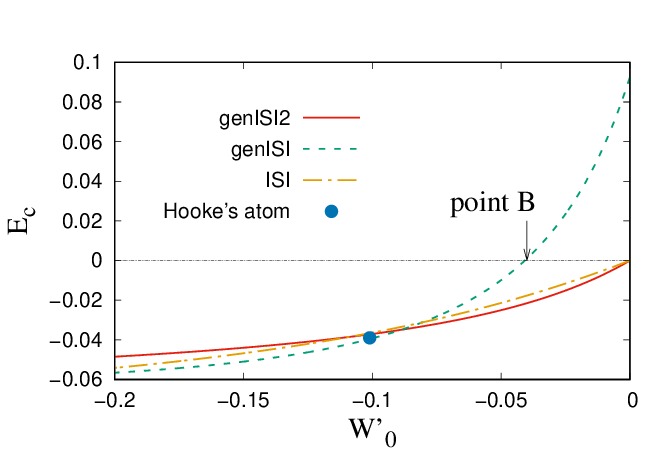}
\includegraphics[width=\columnwidth]{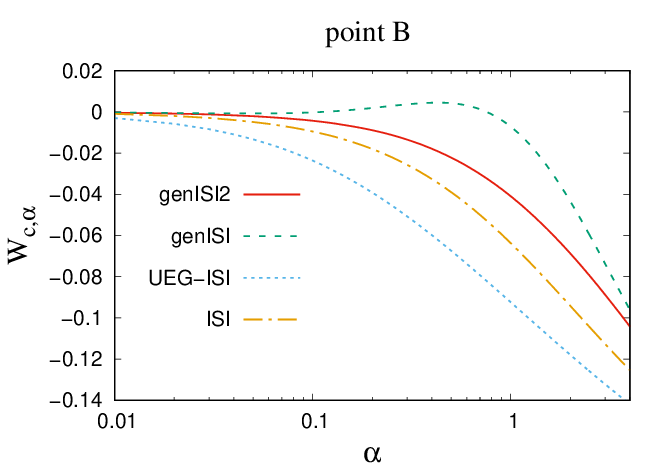}
\caption{
Upper panel: The correlation energy $E_c$ (a.u.) versus $W'_0$ (a.u.) for Harmonium ($\kappa=1/4$), for three ACII functionals. Other ingredients are from Tab. \ref{ta1}.
For the Harmonium ($\kappa=1/4$), $W'_0=-0.101$.\newline
Lower panel: The adiabatic connection correlation integrand $W_{c,\alpha}$ versus $\alpha$ (in log scale), for the point B of the first panel (indicated with an arrow), where $W'_0=-0.04$ Ha.
}
\label{f1}
\end{figure}
%
 To visualize the negativity condition of the correlation energy, let us fix  $W_0$, $W_\infty$ and $W'_\infty$ with the Harmonium values (see Tab. \ref{ta1}) and we vary $W'_0$ (note that for the Harmonium $W'_0=-0.101$). The results are reported in the upper panel of Fig. \ref{f1}, where we observe that genISI correlation energy becomes positive when $W'_0\rightarrow 0$, and genISI2 solves this failure. Moreover, as shown in the lower panel of Fig. \ref{f1}, the correlation integrand $W_{c,\alpha}$ of genISI2 remains smooth, with the correct slope at $\alpha\rightarrow 0$, even for the difficult case when $W'_0$ is small ($W'_0=-0.04$), where genISI curve becomes wrongly positive and non-monotonic.

 \twocolumngrid
%
 

\end{document}